\documentclass{article}

\usepackage{arxiv}

\usepackage{adjustbox}
\usepackage{amsfonts}
\usepackage[fleqn]{amsmath}
\usepackage{amssymb}
\usepackage{amsthm}

\usepackage{bm}

\usepackage{booktabs}
\usepackage{cases}
\usepackage{cite}  
\usepackage{color}
\usepackage{enumerate}
\usepackage{enumitem}
\usepackage{fleqn}
\usepackage{gnuplot-lua-tikz}
\usepackage{latexsym}
\usepackage{mathrsfs}
\usepackage{newtxtext}
\usepackage{pifont}
\usepackage{mathtools}
\usepackage{multirow}
\usepackage[caption=false,font=footnotesize]{subfig}
\usepackage{tikz}
\usetikzlibrary{backgrounds, fit, shapes, calc, positioning}
\usepackage{txfonts}
\usepackage{url}
\usepackage{xcolor}

\usepackage[linesnumbered,ruled,vlined]{algorithm2e}
\SetKwRepeat{Do}{do}{while}
\DontPrintSemicolon

\SetCommentSty{mycommfont}

\makeatletter
\newcommand{\removelatexerror}{\let\@latex@error\@gobble}
\makeatother

\theoremstyle{definition}

\newtheorem{problem}{Problem}

\theoremstyle{remark}

\definecolor{darkgreen}{rgb}{0.1,0.6,0.1}

\newcommand{\footurl}[1]{\footnote{\footnotesize \url{#1}}}

\title{International Competition on Graph Counting Algorithms 2023
}

\newif\ifuniqueAffiliation
\uniqueAffiliationtrue

\usepackage{authblk}

\newbox{\orcid}\sbox{\orcid}{\includegraphics[scale=0.06]{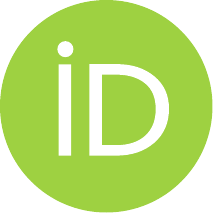}} 
\author[1,2]{%
  {Takeru Inoue\thanks{\texttt{takeru.inoue@ieee.org}}}%
}
\author[1]{%
  {Norihito Yasuda}
}
\author[3]{%
  {Hidetomo Nabeshima}
}
\author[1]{%
  {Masaaki Nishino}
}
\author[1]{%
  {Shuhei Denzumi}
}
\author[4]{%
  {Shin-ichi Minato}
}
\affil[1]{NTT Communication Science Laboratories, NTT Corporation, Kyoto, Japan}
\affil[2]{NTT Network Innovation Laboratories, NTT Corporation, Kanagawa, Japan}
\affil[3]{Graduate Faculty of Interdisciplinary Research, University of Yamanashi, Yamanashi, Japan}
\affil[4]{Graduate School of Informatics, Kyoto University, Kyoto, Japan}

\renewcommand{\shorttitle}{\textit{arXiv} Template}

\hypersetup{
pdftitle={International Competition on Graph Counting Algorithms 2023
},
pdfsubject={cs.DS},
pdfauthor={Takeru Inoue, Norihito Yasuda, Hidetomo Nabeshima, Masaaki Nishino, Shuhei Denzumi, Shin-ichi Minato},
pdfkeywords={Graph counting algorithms, algorithm competitions, and ICGCA

},
}

\begin{document}
\maketitle

\begin{abstract}
This paper reports on the details of the International Competition on Graph Counting Algorithms (ICGCA) held in 2023.
The graph counting problem is to count the subgraphs satisfying specified constraints on a given graph.
The problem belongs to \#P-complete, a computationally tough class.
Since many essential systems in modern society, e.g., infrastructure networks, are often represented as graphs, graph counting algorithms are a key technology to efficiently scan all the subgraphs representing the feasible states of the system.
In the ICGCA, contestants were asked to count the \emph{paths} on a graph under a length constraint.
The benchmark set included 150 challenging instances, emphasizing graphs resembling infrastructure networks.
Eleven solvers were submitted and ranked by the number of benchmarks correctly solved within a time limit.
The winning solver, \texttt{TLDC}, was designed based on three fundamental approaches: backtracking search, dynamic programming, and model counting or \#SAT (a counting version of Boolean satisfiability).
Detailed analyses show that each approach has its own strengths, and one approach is unlikely to dominate the others.
The codes and papers of the participating solvers are available: \url{https://afsa.jp/icgca/}.


\end{abstract}


\section{Introduction}
\label{sec:intro}

Complex modern societal systems, composed of various components like infrastructure networks, are often represented as graphs to capture their underlying structures.
These systems are usually operated or controlled to satisfy predefined constraints specific to each system, in order to function effectively.
The constraints are often represented as a set of feasible \emph{subgraphs} on the graph representation of the system; e.g., for an infrastructure network to function properly, operators represent the network topology as a graph and ensure the existence of a path, a subgraph connecting a supply source and a consumption point. 
Examining feasible system states, assessing their probabilities, and searching for the optimal state require comprehensive management of the set of feasible subgraphs.
As a result, the problem of counting feasible subgraphs, known as the graph counting (GC) problem, emerges as the most fundamental form of algorithms that efficiently scan all the feasible subgraphs~\cite{minato17}.
Recently, GC algorithms and their variants have found practical applications, especially in domains like infrastructure networks, which require examining every state without overlooking them~\cite{xing15}.
Examples of applications include infrastructure networks such as power~\cite{inoue14,duenas-osorio17}, communications~\cite{nakamura23fast}, and railroads~\cite{kawahara17solving}; floor planning and disaster evacuation~\cite{takizawa15,takizawa13}; propagation phenomena in social networks~\cite{maehara17} and epidemics~\cite{ishioka19}; internal networks of supercomputers~\cite{koibuchi16}; and institutional design such as electoral districting~\cite{kawahara17generating}. 
To address the growing demand for GC algorithms and to further foster them, the International Competition on Graph Counting Algorithms (ICGCA) was held in 2023.

\subsection{Literature Survey on Graph Counting Algorithms}
\label{sec:intro-survey}

The GC problem is known to be \#P-complete, a computationally tough class~\cite{valiant79}.
Since the number of paths can be exponential with respect to the graph size, efficient algorithms have been enthusiastically studied for decades.
This paper highlights the following three approaches to GC algorithm research. 

\begin{itemize}
\item Backtracking (BT):
Initial research on GC, which emerged in the 1960s and 1970s, relied on pure algorithmic approaches that visit and output feasible subgraphs one at a time, using search methods such as \emph{backtracking}~\cite{tarjan73,read75}.
This approach is relevant to the limited memory capacity available at the time and inherently requires time proportional to the number of feasible subgraphs.
Thus, it does not scale well for instances with a large number of subgraphs.

\item Dynamic programming (DP):
Since the end of the 20th century, efficient counting methods using \emph{dynamic programming} on decision diagrams have been intensively studied~\cite{sekine95,coudert97}.
These methods successfully handle an astronomical number of subgraphs in a graph with hundreds of edges.
In the early stages of DP research, algorithms were developed for specific problems like network reliability evaluation~\cite{hardy07} and path enumeration~\cite{knuth11}.
Common features among these algorithms were identified, and a versatile algorithmic framework called ``frontier-based search'' (FBS)~\cite{kawahara17frontier} was developed, along with publicly released implementations such as TdZdd~\footurl{https://github.com/kunisura/TdZdd/} and Graphillion~\footurl{https://github.com/takemaru/graphillion/}\cite{inoue16graphillion}.
Thanks to their scalability and generality, these implementations have been utilized for many of the above applications.

\item Model Counting (MC):
In the field of artificial intelligence (AI), \emph{model counting} is a fundamental problem encompassing graph counting~\cite{gomes21}.
MC is sometimes referred to as \#SAT due to its counting version of the satisfiability (SAT) problem. 
In MC, systems are represented as Boolean formulae, and the goal is to count the feasible models that satisfy the given constraints.
MC algorithms can be applied to GC by representing graphs and their constraints as Boolean formulae.
Since interest has been rising in counting the number of models for practical problems, the MC competition\footurl{https://mccompetition.org/}~\cite{fichte20} was launched in 2020. 
Note that since MC involves several techniques, we indicate which one is employed in each solver.
\end{itemize}


\subsection{Related Algorithm Competitions and ICGCA Goals}
\label{sec:intro-competition}

Competitions serve various purposes, from specific problems for industry needs to fundamental advancements for academic research.
In the former case, competitions sometimes provide large cash awards like the Netflix Prize~\cite{bell07}.
On the other hand, mathematical problem solving like GC usually follows the latter purpose and often lacks specific monetary rewards.
We follow a direction in the community of mathematical problem solving where many competitions and challenges have already been organized: SAT\footurl{http://www.satcompetition.org/}~\cite{froleyks21} (22 editions), SMT\footurl{https://smt-comp.github.io/2023/}~\cite{weber19} (19 editions), MaxSAT\footurl{https://maxsat-evaluations.github.io/2023/}~\cite{li21} (18 editions), CSP\footurl{https://www.minizinc.org/challenge.html} (16 editions), QBF\footurl{http://www.qbflib.org/index_eval.php}~\cite{marin16} (13 editions), DIMACS\footurl{http://dimacs.rutgers.edu/programs/challenge/}~\cite{demetrescu09} (12 editions), PACE\footurl{https://pacechallenge.org/}~\cite{grossmann22} (8 editions), MC~\cite{fichte20} (3 editions), and CoRe\footurl{https://core-challenge.github.io/2023/} (2 editions).

Despite the success of MC competition in addressing more general problems than GC, we organized a GC competition for the following reasons.
GC algorithms have features distinct from MC algorithms, enhancing computational efficiency greatly.
This is implicitly supported by the fact that GC is used instead of MC in many application areas.
On the other hand, GC and MC (and BT) share several technical features, and they could improve each other's performance through positive interactions.
Unfortunately, these approaches have been pursued relatively independently and have had limited mutual influence.


ICGCA's primary goal is to bring the three approaches (BT, DP, and MC) onto the same platform, clarify their strengths and weaknesses, and foster the development of better algorithms through their interaction.
The competition was carefully designed to prevent any one approach from gaining an undue advantage.
The ICGCA also aims to identify new challenging benchmarks and promote novel approaches.
We hope that active participation, collaboration, and long-term improvements will broaden the applicability of GC algorithms in practice and inspire many new applications.


The rest of this paper is organized as follows.
Section~\ref{sec:overview} overviews the ICGCA and Section~\ref{sec:benchmark} describes the benchmark set.
Section~\ref{sec:solver} presents the submitted solvers and their results, which are analyzed in detail in Section~\ref{sec:analysis}.
The paper is concluded in Section~\ref{sec:conclusion}.

\section{ICGCA Overview}
\label{sec:overview}

This section provides an overview of ICGCA, starting with the following timeline.
The ICGCA website was opened in April 2023.
Contestants received the benchmark set when they registered on the ICGCA website.
They developed their solvers and submitted them to the ICGCA organizers by July 17, 2023.
The solvers were evaluated by the organizers, and the results were announced on September 7, 2023.

Section~\ref{sec:overview-problem} defines the problem.
Section~\ref{sec:overview-score} describes the scoring method, and Section~\ref{sec:overview-evaluation} shows the evaluation environments.
Note that we have tried to keep the competition simple since this is the first competition on GC algorithms.

\subsection{Problem}
\label{sec:overview-problem}

The ICGCA presented two variants of \emph{path counting} problems.
One specifies a pair of path terminals.
\begin{problem}[PCS: Path Counting for a Single pair]
Given an undirected simple graph $G = (V, E)$, a pair of vertices $\{s, t\} \in V \times V$, and a maximum path length $\ell$, count the exact number of simple paths between $s$ and $t$ whose length is at most $\ell$.
\end{problem}
Note that the path length is defined as the number of edges on the path.

The other variant specifies no vertex pairs and counts the paths between \emph{all} the vertex pairs.
\begin{problem}[PCA: Path Counting for All pairs]
Given an undirected simple graph $G = (V, E)$ and a maximum path length $\ell$, count the exact number of simple paths between all $\{s, t\} \in V \times V$, where $s \neq t$ and the path length is at most $\ell$.
\end{problem}

\begin{figure}[t]
  \centering
  \subfloat[An instance.]{\includegraphics[width=.14\textwidth]{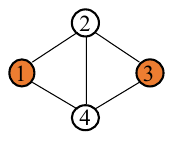}}
  \subfloat[The answer is the two paths with check marks.]{\includegraphics[width=.35\textwidth]{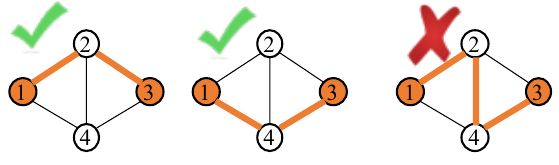}}
  \caption{Example of PCS instance:
    terminal pair is 1 and 3, and maximum path length is two edges.}
  \label{fig:problem}
\end{figure}

Figure~\ref{fig:problem} shows an example of a PCS instance.
A terminal pair (vertices 1 and 3) and a maximum path length of two are given.
Since there are two paths between the terminal pair with at most two edges, the answer is \emph{two} paths.
For a PCA instance (if no terminal pair were specified), the answer is 13 paths.

We focus on \emph{undirected} graphs in this competition, because infrastructure networks, a primary application of GC algorithms, are often modeled as either undirected or bidirectional graphs.
In addition, past work has mainly focused on undirected graphs, and competitions for directed graphs are considered premature.

We also concentrate on \emph{paths} as a subgraph structure to be counted.
Many other options are possible: cycles, trees, forests (sets of trees), matchings (independent vertex/edge sets), cliques, connected components, and partitions.
However, paths are often examined in practical problems, especially for connecting supply sources and consumers in infrastructure networks.
In addition, path counting remains an active research area~\cite{punzi23}, suggesting substantial room for improvement.

We impose a single constraint: \emph{maximum path length}.
Although we can devise various constraints, e.g., Hamiltonicity and way-pointing, too many constraints could overwhelm the contestants.
On the other hand, without any constraints, the amount of feasible paths might also become overwhelming, disadvantaging the BT approach whose computational complexity strongly depends on path counts.
In practical scenarios, it is often acceptable to ignore large detour paths, and so we chose the maximum path length as an essential constraint.


\subsection{Scoring methods}
\label{sec:overview-score}

The winner is the solver that provides correct answers for the largest number of benchmarks. 
Note that each benchmark is allocated a time budget of 600 seconds in wallclock time.
The benchmark set includes 150 instances: 100 for PCS and 50 for PCA.

\#P problems, including GC, pose a severe challenge to scoring methods;
while NP problems can be verified in polynomial time, for \#P problems, an answer's correctness can only be verified by solving the instance itself. 
Although excluding hard instances from the benchmark set could resolve the verification issue, contestants would not be encouraged to develop a solver for hard instances.
The MC competition, which faced the same issue, treated any answer for unknown benchmarks as ``correct''~\cite{fichte20}.
The ICGCA employs a fairer scoring method.
Since wrong answers for benchmarks with many paths are unlikely to match by chance, answers are labeled as ``correct'' if they match among multiple solvers. 
If the answers do not match (either the benchmark was solved by a single solver or the answers were split among solvers), they are labeled ``tentatively correct.''
Then, the scores are presented with a range: [``\# of correct answers,'' ``\# including tentatively correct answers''].
Therefore, there was a risk that the rankings might not be uniquely fixed, but the actual competition resulted in a uniquely fixed order.
Since the answer verification is imperfect as described, solvers with incorrect answers were not penalized.

The ICGCA does not employ the PAR-2 system, a well-known scoring method used in the SAT competition~\cite{froleyks21}, which reflects the time required for answering each benchmark in the rankings.
This is for keeping the competition simple, but the PAR-2 system will be used to analyze competition results in Section~\ref{sec:analysis}.


\subsection{Solver Requirements and Evaluation Environments}
\label{sec:overview-evaluation}

\begin{figure}[t]
  \centering
  \includegraphics[width=.15\textwidth]{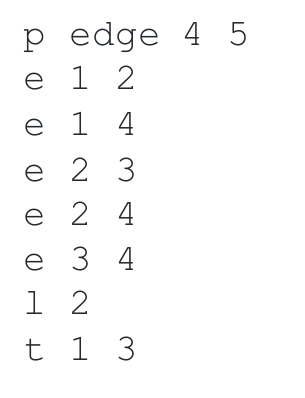}
  \caption{Input for Fig.~\ref{fig:problem} in extended DIMACS file format.}
  \label{fig:input}
\end{figure}

An instance is given in an extended DIMACS file format.
Each line begins with a letter that defines the rest of the line.
The allowed lines are as follows.
\begin{itemize}
\item \texttt{p}: The problem description must have form \texttt{p edge n m}, where \texttt{n} is the number of nodes (numbered 1..n) and \texttt{m} is the number of edges.
\item \texttt{e}: An edge must have form \texttt{e v1 v2}, where \texttt{v1} and \texttt{v2} are the endpoints of the edge.
\item \texttt{l}: The maximum path length must have form \texttt{l len}, where \texttt{len} is the maximum path length. 
\item \texttt{t}: The pair of path terminals must be of form \texttt{t v1 v2}, where \texttt{v1} and \texttt{v2} are the path terminals. This line is omitted for PCA instances.
\end{itemize}
Fig.~\ref{fig:input} shows the input for Fig.~\ref{fig:problem}.
Solvers are required to output just a number of paths to \texttt{stdout}.
The solver codes are made public after the competition.

Solvers were evaluated on an ICGCA organizer computer with 12 CPU cores (Intel Core i7-1260P, up to 4.70 GHz) and 64 GB RAM, running Ubuntu Server 22.04 LTS.
They were allowed to use multiple CPU cores, and portfolio solvers implementing multiple algorithms were also acceptable.
Although the SAT competition had separate tracks for sequential, parallel, and portfolio solvers, the ICGCA had only a single track for parallel and portfolio solvers.
This decision was made to avoid excessively restricting contestants' strategies without hosting multiple tracks.

To ensure correct scoring, we conducted the following evaluation.
First, the pool of solvers was narrowed down to the top three, which were then run through multiple rounds of evaluation on different computers to validate the consistency of their results.


\section{Benchmarks}
\label{sec:benchmark}

Since no benchmark set had been developed for GC competitions, we first established guidelines for preparing one.
\begin{itemize}
\item Focus on real-world graphs and their close counterparts, emphasizing infrastructure networks, i.e., graphs with at most a few thousand edges and sparse connectivity.
\item Avoid biasing any particular approach by conducting preliminary experiments.
\item Cover a spectrum of complexity from simple to challenging instances (those can not be easily solved by existing solvers, but can be solved with new innovations).
\end{itemize}
The preliminary experiments are described in Section~\ref{sec:benchmark-experiment}.
The selection of benchmarks and their statistics is presented in Section~\ref{sec:benchmark-stats}.

\subsection{Preliminary Experiments}
\label{sec:benchmark-experiment}

To measure the hardness of various instances, we conducted preliminary experiments.
We prepared test solvers for the BT and DP approaches, although we found no MC solver that is easily applicable to GC instances.
The BT test solver implements Yen's $k$-shortest path algorithm\footurl{https://github.com/yan-qi/k-shortest-paths-cpp-version/}~\cite{martins03}, which visits paths in ascending order of length until the maximum path length $\ell$ is reached. 
The DP test solver employs Graphillion~\cite{inoue16graphillion}, which processes edges in the order of the path decomposition~\cite{inoue16acceleration}.
Note that these test solvers are easy to use with reasonable computational efficiency, although they are not state-of-the-art.

We prepared 1360 test instances.
The graphs were chosen from typical structures: complete graphs, 2D grid graphs, path-like graphs, and tree-like graphs.
A path-like graph connects small cliques with a few edges in a path-like manner.
A tree-like graph is ditto.
Path-/tree-like graphs mimic infrastructure networks, where regional subnetworks are connected by long-distance links.
Note that although complete graphs are quite different from infrastructure networks, we included a few to understand the performance of the test solvers.
The maximum path length $\ell$ was randomly chosen from the range of the graph diameter to the number of vertices (smaller values were more likely to be chosen).
For PCS instances, the most distant vertex pair, $\{s, t\}$, was selected.
The computer used for the preliminary experiments had 32 GB of memory, less than for real environments.
The test solvers were run in a single thread.

\begin{table}[t] 
  \caption{Results of preliminary experiments}
  \label{table:experiment}
  \begin{center}
    \begin{tabular}{l|r|r}
      \toprule
                     & Solved by BT & Unsolved by BT \\
      \midrule
      Solved by DP   & 392          & 283            \\
      Unsolved by DP & 112          & 573            \\
      \bottomrule  
    \end{tabular} 
  \end{center} 
\end{table}

Table~\ref{table:experiment} shows the results of the preliminary experiments.
392 instances were solved by both the BT and DP test solvers; 573 instances were not solved by either one.
Overall, the BT test solver was impacted by the maximum path length, because a larger maximum path length yields more paths.
The DP test solver was affected by \emph{pathwidth}, which is a graph parameter that indicates the dissimilarity to a path: one for a path graph and $n - 1$ for a complete graph with $n = |V|$. 
The strong dependence of the DP approach on the pathwidth was known from the literature~\cite{inoue16acceleration}.


The results were used to build a machine learning model, which was used to select instances with diverse hardness levels without favoring one approach over another.
The feature space consists of the pathwidth, $\ell$, $n$, and the graph's mean degree.
The model was built using the support vector machine.

\subsection{Benchmark Selection and Statistics}
\label{sec:benchmark-stats}

\begin{table}[t] 
  \caption{Number of benchmarks with real and synthesized graphs for PCS and PCA}
  \label{table:benchmark}
  \begin{center}
    \begin{tabular}{l|r|r}
      \toprule
          & Real graphs & Synthesized graphs \\
      \midrule
      PCS & 15          & 85                 \\
      PCA & 30          & 20                 \\
      \bottomrule  
    \end{tabular} 
  \end{center} 
\end{table}

We selected benchmarks from real and synthetic graphs and show the number of selected benchmarks with real and synthetic graphs in Table~\ref{table:benchmark}.

First, we describe the real graphs included in the benchmark set.
We collected 45 graphs, which are sparse with at most a few thousand edges, from communication networks (TopologyZoo\footurl{http://topology-zoo.org/}~\cite{knight11}, RocketFuel\footurl{https://research.cs.washington.edu/networking/rocketfuel/}~\cite{spring02}, SNDlib\footurl{http://sndlib.zib.de/home.action}, and JPN48\footurl{https://www.ieice.org/cs/pn/jpn/jpnm.html}) and software engineering domains (Rome\footurl{http://www.graphdrawing.org/data.html}).
These graphs have been used in recent studies of GC algorithms~\cite{suzuki20,nakamura21}.
The 45 graphs were associated with each benchmark, resulting in 45 benchmarks.
No selection based on the hardness model was made for real graphs to include as many of them as possible in the benchmark set.

The remaining 105 benchmarks were created with synthetic graphs.
To increase the diversity of those used in the preliminary experiments, we randomly modified them by removing three or fewer edges from the graph and rewiring 25 or fewer edges.
105 synthetic benchmarks were selected based on the hardness model.
59 benchmarks were predicted as unlikely to be solved by either the DP or BT test solvers, which implies they are quite challenging.
73 benchmarks were unlikely to be solved by the DP test solver, and 79 were unlikely to be solved by the BT test solver: roughly the same number.
%

\begin{figure}[t]
  \centering
  \includegraphics[width=.5\textwidth]{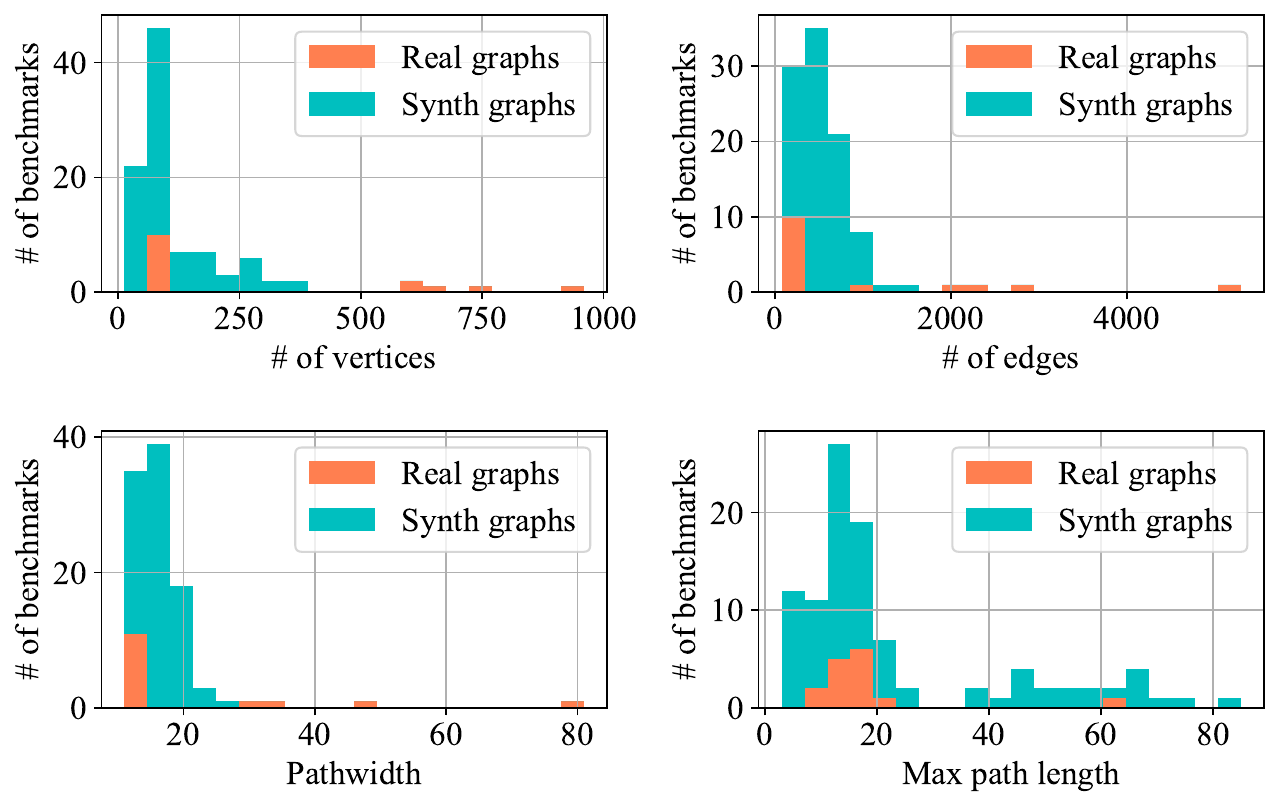}
  \caption{Histograms for instance parameters in PCS benchmark set.}
  \label{fig:benchmark-pcs}
\end{figure}

\begin{figure}[t]
  \centering
  \includegraphics[width=.5\textwidth]{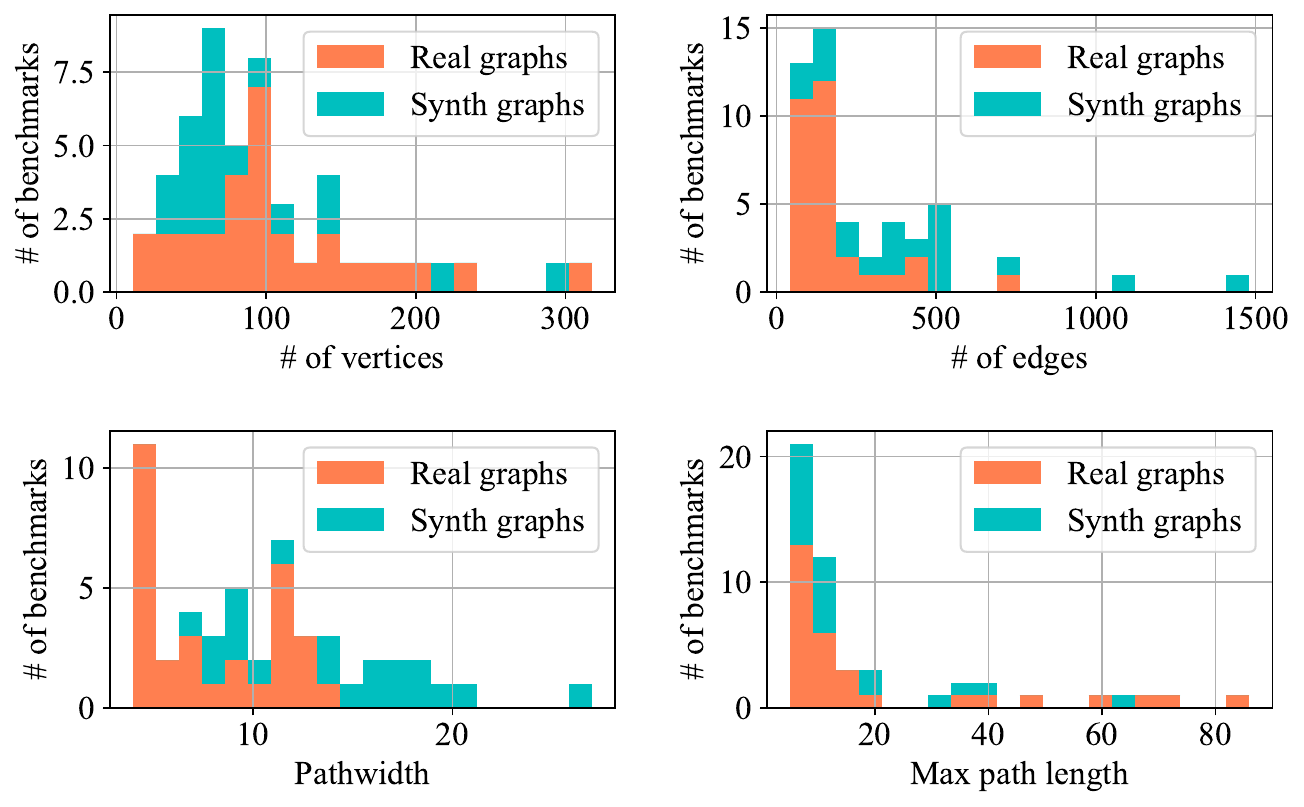}
  \caption{Histograms for instance parameters in PCA benchmark set.}
  \label{fig:benchmark-pca}
\end{figure}

\begin{figure}[t]
  \centering
  \includegraphics[width=.45\textwidth]{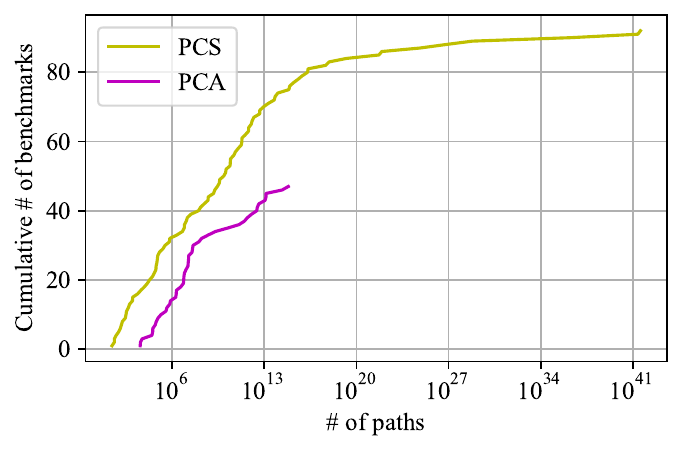}
  \caption{Cumulative path counts.}
  \label{fig:cdf-counts}
\end{figure}

Next we examine the statistics of the benchmark set.
Figs.~\ref{fig:benchmark-pcs} and~\ref{fig:benchmark-pca} show the parameter distributions for the PCS and PCA benchmark sets.
The PCA benchmarks, which count the paths for all the vertex pairs, have smaller parameters than the PCS benchmarks.
Although all the parameters are mainly distributed around smaller regions, we found some very challenging benchmarks in the larger regions.
In particular, some PCS benchmarks with real graphs have extremely large parameters, because the hardness is not controlled for real graphs.
Fig.~\ref{fig:cdf-counts} shows the cumulative path counts for the 139 benchmarks that were solved by at least one solver.
The PCS benchmarks have many more paths, $10^{41}$ paths at most, due to the larger parameters.
The PCA benchmarks have fewer paths, even though the paths for all the vertex pairs are counted.

\section{Solvers and Results}
\label{sec:solver}

Section~\ref{sec:solver-results} presents the ICGCA result, and Section~\ref{sec:solver-solver} overviews the participating solvers.

\subsection{Results}
\label{sec:solver-results}

\begin{table}[t]
  \caption{Solver scores}
  \label{table:score}
  \begin{center}
    \begin{tabular}{l|r|r|r}
      \toprule
                         & Overall    & PCS      & PCA      \\
      \midrule
      \texttt{TLDC}      & [135, 136] & [90, 91] & 45       \\
      \texttt{diodrm}    & [131, 134] & [86, 87] & [45, 47] \\
      \texttt{TAG}       & 126        & 83       & 43       \\
      \texttt{Drifters}  & 110        & 71       & 39       \\
      \texttt{NaPS+GPMC} & 105        & 72       & 33       \\
      \texttt{KUT\_KMHT} & 99         & 61       & 38       \\
      \texttt{asprune}   & 65         & 40       & 25       \\
      \texttt{castella}  & 39         & 29       & 10       \\
      \texttt{dimitri}   & [11, 12]   & 10       & [1, 2]   \\
      \texttt{Cypher}    & 4          & 2        & 2        \\
      \texttt{PCSSPC}    & 0          & 0        & 0        \\
      \bottomrule
    \end{tabular}
  \end{center}
\end{table}

The ICGCA received 11 solvers from four countries: Austria, Germany, Japan, and USA.
Table~\ref{table:score} ranks them.
As noted in Section~\ref{sec:overview-score}, the scores (the number of correctly solved benchmarks) are denoted in square brackets if they have a range.
The best-performing solver overall on the combination of PCS and PCA benchmarks is \texttt{TLDC}, and the runner-up is \texttt{diodrm}; \texttt{diodrm} solved the most instances for PCA.
Unfortunately, the bottom three solvers, \texttt{dimitri}, \texttt{Cypher}, and \texttt{PCSSPC}, did not perform as expected; they crashed during execution or yielded more wrong answers than correct ones.
Therefore, we excluded them from the following analyses.

\begin{figure}[t]
  \centering
  \includegraphics[width=.45\textwidth]{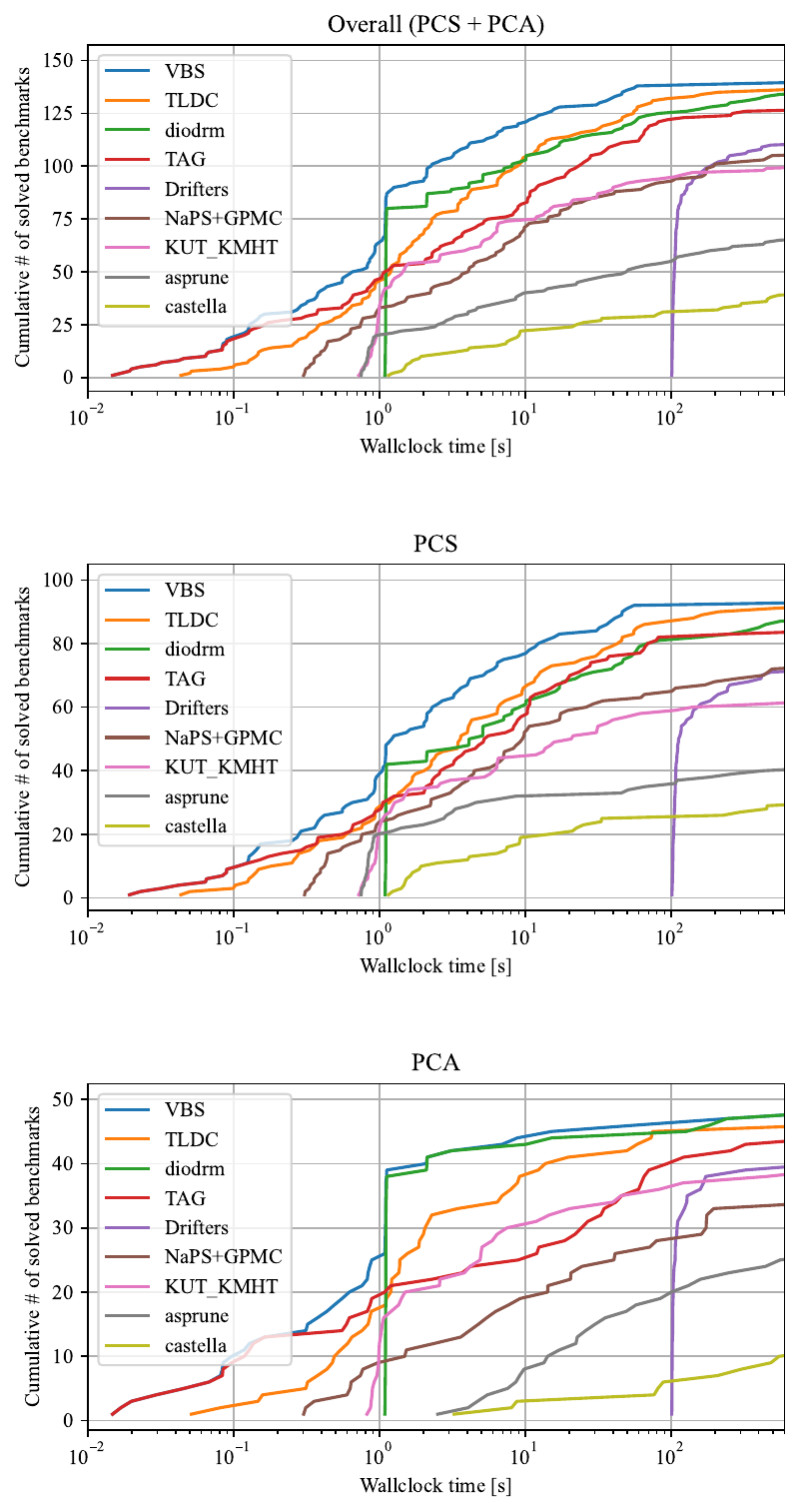}
  \caption{Performance of top eight solvers and VBS.}
  \label{fig:cdf-time}
\end{figure}

Figure~\ref{fig:cdf-time} shows the cumulative solved benchmark plot together with the Virtual Best Solver (VBS).
The VBS is a fictitious solver consisting of all the solvers that actually participated in the competition and an oracle which, given an input instance, invokes the solver that performed the best on that instance.
The VBS performance highlights a certain upper bound on the performance achievable by the participating solvers.
For the PCS benchmark set, \texttt{TLDC} maintained the lead from three seconds onward.
For the PCA benchmark set, \texttt{diodrm} showed surprising performance comparable to the VBS, demonstrating its overwhelming strength in PCA.
\texttt{TAG} showed great start-up and solved more instances in the sub-second range.
Note that the steep increase for \texttt{diodrm} and \texttt{Drifters} at 1 and 100 seconds reflects the fixed amount of preprocessing time spent on them.

\subsection{Solvers}
\label{sec:solver-solver}

\begin{table}[t]
  \caption{Solver approaches}
  \label{table:approach}
  \begin{center}
    \begin{tabular}{l|l|l|l|l|l}
      \toprule
                         & Portfolio  & BT         & DP                    & MC (via)         & Other      \\
      \midrule
      \texttt{TLDC}      & \checkmark & \checkmark & \checkmark            & \checkmark (ASP) &            \\
      \texttt{diodrm}    & \checkmark & \checkmark & \checkmark            &                  &            \\
      \texttt{TAG}       & \checkmark &            & \checkmark \checkmark &                  &            \\
      \texttt{Drifters}  &            &            & \checkmark            &                  &            \\
      \texttt{NaPS+GPMC} &            &            &                       & \checkmark (PB)  &            \\ 
      \texttt{KUT\_KMHT} & \checkmark &            & \checkmark \checkmark &                  & \checkmark \\
      \texttt{asprune}   &            &            &                       & \checkmark (ASP) &            \\
      \texttt{castella}  &            &            &                       & \checkmark (CSP) &            \\
      \texttt{dimitri}   &            &            &                       &                  & \checkmark \\
      \texttt{Cypher}    &            &            & \checkmark            &                  &            \\
      \texttt{PCSSPC}    &            & \checkmark &                       &                  &            \\
      \bottomrule
    \end{tabular}
  \end{center}
\end{table}

This subsection overviews the top eight solvers.
The following description is based on short papers submitted by the contestants and our glances at their code.
The codes and papers for all the solvers are available online: \url{https://afsa.jp/icgca/}.
The approaches employed by each solver are shown in Table~\ref{table:approach}.

\begin{itemize}
\item
\texttt{TLDC} (Too Long; Didn't Count) is a portfolio solver consisting of all three approaches: BT, DP, and MC.
This solver performs BT in a depth-first search (DFS) manner, but it is much more efficient than the traditional naive implementation thanks to the techniques described later.
It also employs FBS, a DP method~\cite{kawahara17frontier}, with an edge processing order based on the tree decomposition computed by htd\footurl{https://github.com/mabseher/htd/}~\cite{abseher17}.
This solver has several ideas that make the solver more efficient, which are elegantly applied to both BT and DP methods, e.g., the search space is pruned based on the path length, and graph isomorphism tested by nauty\footurl{https://pallini.di.uniroma1.it/}~\cite{mckay14} reduces the cache entries.
In addition, TLDC uses MC-derived ideas, e.g., it employs an ASP solver named Clingo\footurl{https://potassco.org/clingo/} to check whether the maximum length of a path can be achieved.
\texttt{TLDC} is well-designed based on all three approaches and won the competition.
\item
\texttt{diodrm} is a portfolio solver following the BT and DP approaches.
Like \texttt{TLDC}, this solver also employs DFS and FBS, but the FBS's edge order is determined by the community-based order.
A large instance is partitioned by a method called \emph{meet in the middle} to facilitate precomputation and parallelization.
\texttt{diodrm} overwhelmed the other solvers for PCA by introducing the following innovative DFS techniques.
A search is performed not for each vertex pair but only from each vertex, which also allows easy parallelization.
The search is further accelerated by exploiting the fact that, for a graph with small neighborhood diversity, a path is still formed even if the vertices in an adjacent set are swapped.
\item
\texttt{TAG} is a portfolio solver consisting of two DP methods: FBS and HAMDP (Hamiltonian DP).
FBS introduces a pruning strategy with path length and vertex degree, based on a traditional FBS design (implemented with TdZdd and the edge order is determined based on path decomposition).
HAMDP is an interesting extension of a Hamiltonian-paths algorithm and outperforms FBS on some PCA benchmarks, according to their paper.
\item
\texttt{Drifters} implements FBS with TdZdd and utilizes Chokudai search to perform path decomposition to determine the edge processing order.
\item
\texttt{NaPS+GPMC} is an MC solver designed for path counting problems.
First, it produces Pseudo-Boolean (PB) constraints of a path, which are converted into CNF form using NaPS\footurl{https://www.trs.css.i.nagoya-u.ac.jp/projects/NaPS/}~\cite{sakai15}.
The number of models (paths) is counted with a variant of the projection model counting solver GPMC\footurl{https://git.trs.css.i.nagoya-u.ac.jp/k-hasimt/GPMC/}.
It was modified for the path counting by removing the time-consuming preprocessing, adopting a new variable choice strategy along with a path, and introducing hierarchical counting strategy.
\item
\texttt{KUT\_KMHT} is a portfolio solver consisting of two DP methods and another type of counting method.
The two DP methods are FBS with TdZDD and a bounding algorithm originally designed for network reliability evaluation~\cite{niu11}.
The other method employs an exact cover algorithm\footurl{https://github.com/aaw/cover/} modified for path counting.
\item
\texttt{asprune} is an MC solver.
First, this solver converts a given instance to ASP facts with an ASP encoding for path counting.
Then, the solver \emph{enumerates} all the paths with Clingo.
This indicates that even an explicit enumeration technique could achieve this level of results.
%
\item
\texttt{castella} is an MC solver.
For a given instance, it generates a CSP, which is translated into a SAT instance using Sugar\footurl{https://cspsat.gitlab.io/sugar/}~\cite{tamura09}.
Then, the solver counts the path with GPMC, as does \texttt{NaPS+GPMC}.
The gap between \texttt{castella} and \texttt{NaPS+GPMC} is due to the latter's several techniques specialized in path counting.
\end{itemize}

The top three solvers, \texttt{TLDC}, \texttt{diodrm}, and \texttt{TAG}, are all portfolio types.
It is interesting that \texttt{Drifters} and \texttt{NaPS+GPMC} implement the different approaches but they performed equally well.
In Section~\ref{sec:analysis-comp}, we compare these two approaches by considering the two solvers as representatives of each approach.
\texttt{KUT\_KMHT} and \texttt{dimitri} implement approaches beyond our expectations, which are categorized as ``others''.
For example, \texttt{dimitri} employ an algebraic approach based on graph isomorphism.

\section{Analyses of Results}
\label{sec:analysis}

This section provides an additional analysis of the results by delving deeper beyond the rankings.
Section~\ref{sec:analysis-type} examines the results by graph type, and Section~\ref{sec:analysis-par2} discusses the PAR-2 scores.
Section~\ref{sec:analysis-vbs} analyzes the contributions of each solver to the VBS.
Section~\ref{sec:analysis-param} investigates the impact of instance parameters, and Section~\ref{sec:analysis-similarity} discusses the similarity of the participating solvers.
Section~\ref{sec:analysis-comp} compares the DP and MC approaches.

\subsection{Scores per Graph Types}
\label{sec:analysis-type}

\begin{figure}[t]
  \centering
  \includegraphics[width=.45\textwidth]{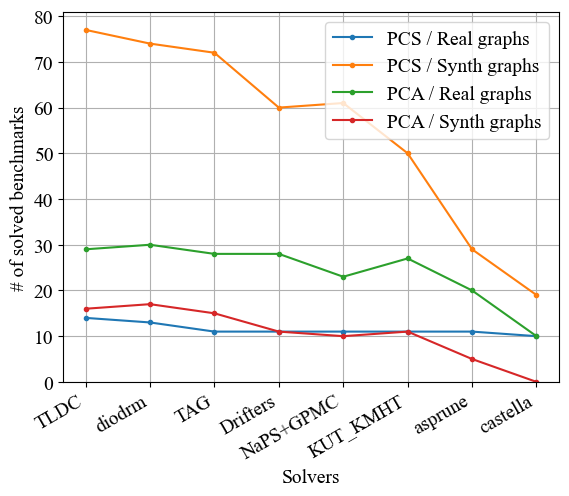}
  \caption{Number of solved benchmarks per graph type.}
  \label{fig:line-type}
\end{figure}

This subsection examines the results for real and synthetic graphs separately (Fig.~\ref{fig:line-type}).
The solvers are ordered on the $x$-axis by their overall ranking.
For all four types, combinations of PCS/PCA and real/synthetic graphs, the lines decline from left to right; the rankings do not change significantly among the types.
For the PCS benchmarks with real graphs, the differences among solvers are small, because many instances are fairly easy, except for a few (Fig.~\ref{fig:benchmark-pcs}).
For the PCA benchmarks with real graphs, \texttt{KUT\_KMHT} performed well compared to the other types. 
Effective techniques might be found for such instances by exploring the unique innovations of \texttt{KUT\_KMHT}.

\subsection{PAR-2 Scores}
\label{sec:analysis-par2}

\begin{figure}[t]
  \centering
  \includegraphics[width=.45\textwidth]{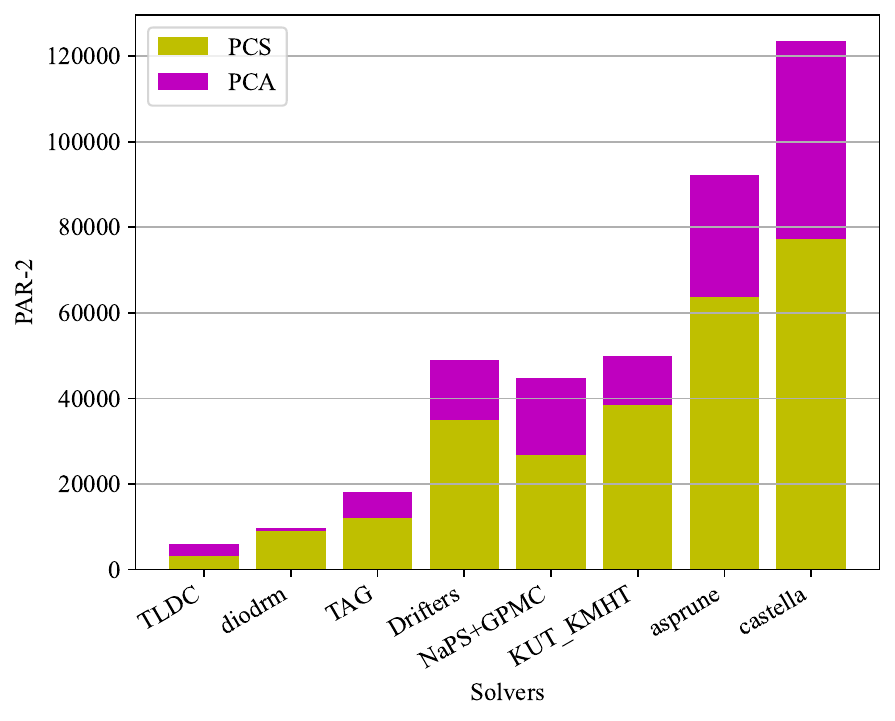}
  \caption{PAR-2 scores (lower is better).}
  \label{fig:bar-par2}
\end{figure}

This subsection ranks the solvers based on the PAR-2 scoring system~\cite{froleyks21}. 
The PAR-2 system assigns as many points as the amount of time (in seconds) required to answer the instance, while it assigns twice the time limit (1,200 points) if the instance was not solved.
This means that the lower a solver's score is, the better its performance.

Figure~\ref{fig:bar-par2} shows the PAR-2 scores accumulated for the 139 benchmarks solved in the competition.
Compared to the rankings by the number of solved benchmarks, \texttt{Drifters}'s ranking worsened because it spent 100 seconds on preprocessing for each benchmark (note that \texttt{Drifters} would not have relied on a fixed-time preprocessing strategy if the ICGCA had employed the PAR-2 system).
Surprisingly, \texttt{diodrm} was not penalized in the least by the PCA benchmarks, demonstrating its overwhelming strength for PCA.
In the following analyses, the computation time for the unsolved benchmarks is considered to be 1,200 seconds according to the PAR-2 system.

\subsection{Contributions to Virtual Best Solver}
\label{sec:analysis-vbs}

The VBS solves all the instances solved by at least one solver with the best time.
By quantifying how much each participating solver contributes to the VBS performance, we can identify which solver is the most important in the state-of-the-art for path counting.
We examine the following three metrics derived from the notion of VBS.
\begin{itemize}
\item VBS-1 ``The fastest takes it all'': For each solver, we count the number of benchmarks in which it was the fastest to solve an instance.
\item VBS-2 ``Time aware, but proportional'': A solver $S$ solving an instance $I$ in time $T_I^S$ obtains $T_I^{VBS} / T_I^S$, where $T_I^{VBS}$ is the computation time of VBS on $I$.
\item VBS-3 ``Split the point for solving'': A solver $S$ solving an instance $I$ obtains $1 / |\mathcal{S}_I|$, where $\mathcal{S}_I$ is the set of solvers solving $I$.
\end{itemize}

\begin{figure}[t]
  \centering
  \includegraphics[width=.45\textwidth]{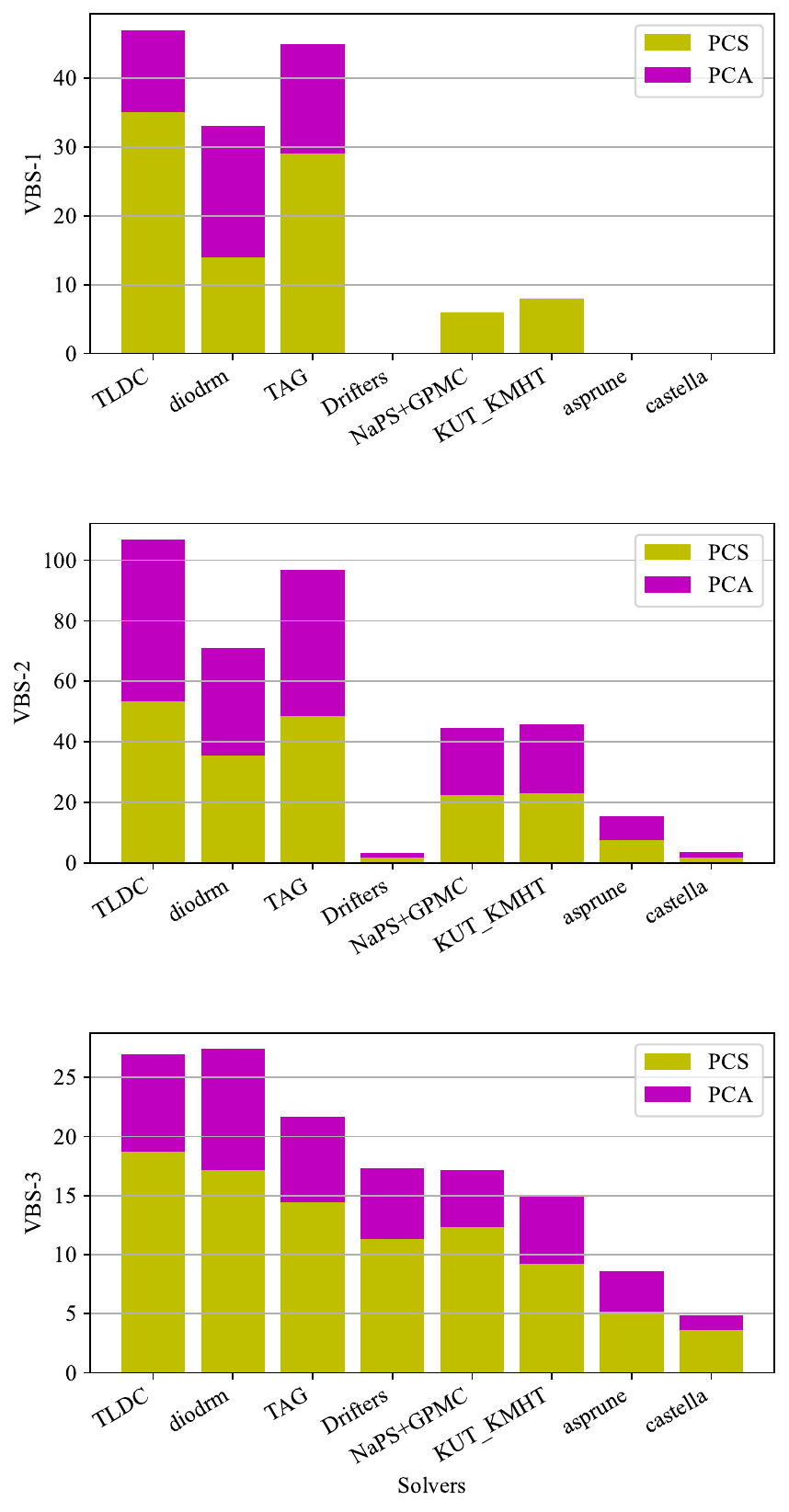}
  \caption{VBS metrics.}
  \label{fig:bars-vbs}
\end{figure}

Figure~\ref{fig:bars-vbs} plots the three metrics.
For VBS-1 and VBS-2, which both involve computation time, \texttt{diodrm} and \texttt{Drifters} with a fixed amount of preprocessing time have lower ranks (nevertheless, \texttt{diodrm} got the most points for PCA in VBS-1).
On the other hand, \texttt{KUT\_KMHT} improved its ranking because it showed comparable performance with the third solver (\texttt{TAG}), up to 10 seconds (Fig.~\ref{fig:cdf-time}).
For VBS-3, \texttt{diodrm} slightly outperforms \texttt{TLDC} because there are three benchmarks solved only by \texttt{diodrm}, while just one benchmark was solved only by \texttt{TLDC}.

\subsection{Impact of Instance Parameters}
\label{sec:analysis-param}

\begin{table}[t] 
  \caption{Correlation between instance parameters and VBS computation time}
  \label{table:corr}
  \begin{center}
    \begin{tabular}{l|r|r|r}
      \toprule
                             & Overall & PCS  & PCA  \\
      \midrule
      \# $n$ of vertices     & 0.22    & 0.23 & 0.16 \\
      \# $m$ of edges        & 0.28    & 0.28 & 0.35 \\
      Pathwidth              & 0.32    & 0.31 & 0.39 \\
      Max path length $\ell$ & 0.47    & 0.54 & 0.29 \\
      \bottomrule  
    \end{tabular} 
  \end{center} 
\end{table}

This subsection examines the impact of the instance parameters on the computation time.
The parameters investigated include the number $n$ of vertices, the number $m$ of edges, the pathwidth, and the maximum path length $\ell$.
VBS's computation is used as a representative for all the solvers.
Table~\ref{table:corr} shows the Pearson correlation coefficients between the parameters and the computation time.
For the PCS benchmark set, the maximum path length has a significant impact, followed by the pathwidth.
This is because pruning strategies based on path length, which are employed by multiple solvers, would work effectively only for short paths.
For the PCA benchmark set, pathwidth has the largest impact, while the maximum path length has a much weaker impact.
This could be because paths between very close vertices are less constrained by the maximum path length.


\subsection{Solver Similarity}
\label{sec:analysis-similarity}

To investigate the similarity among the solvers, we define a similarity metric based on the computation times.
We first removed 11 benchmarks that were not solved by any solver.
For the remaining 139 benchmarks, a PAR-2 score is assigned to each solver-benchmark pair;
i.e., each solver $S$ is thus associated with the PAR-2 scores $S_1$, $\ldots$, $S_{139}$.
The similarity of two solvers, $S$ and $S'$, normalized to the interval $[0, 1]$, is defined as,
\begin{align}
  \text{similarity}(S, S') = 1 - \frac{\sum | S_i - S'_i |}{139 \cdot 1200}.
\end{align}

\begin{figure}[t]
  \centering
  \includegraphics[width=.45\textwidth]{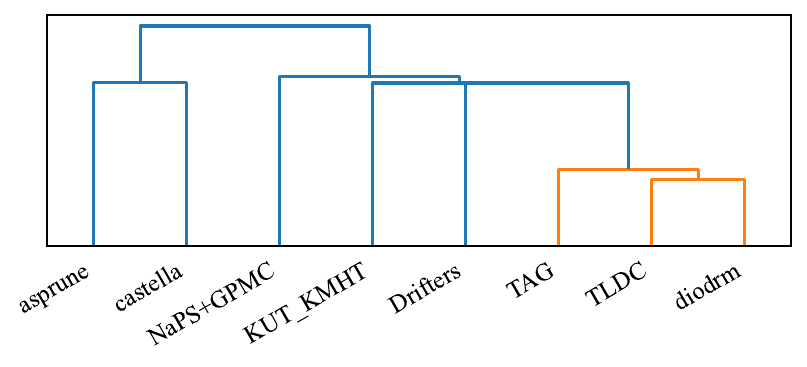}
  \caption{Dendrogram based on computation-time similarity of solvers.}
  \label{fig:dendrogram-similarity}
\end{figure}

The solvers are clustered based on their similarity and are illustrated as a dendrogram (Fig.~\ref{fig:dendrogram-similarity}).
The height at which two solvers or clusters is joined reflects their similarity.
For example, \texttt{TLDC} and \texttt{diodrm} solved most of the benchmarks, so they are joined low in the dendrogram.
The top three solvers (\texttt{diodrm}, \texttt{TLDC}, and \texttt{TAG}) are clustered in the dendrogram.
Interestingly, all the DP solvers (\texttt{diodrm}, \texttt{TLDC}, \texttt{TAG}, \texttt{Drifters}, and \texttt{KUT\_KMHT}) are classified in the same subtree exclusively against the MC solvers, implying that each approach has its own characteristics.



\subsection{Comparison between DP and MC Solvers}
\label{sec:analysis-comp}

This subsection examines the dissimilarity between the DP and MC approaches by considering \texttt{Drifters} and \texttt{NaPS+GPMC} as representatives of both approaches.
\texttt{KUT\_KMHT}, another DP solver that solved roughly an equal number of benchmarks as both solvers, is also employed for comparison.
The three solvers, \texttt{Drifters}, \texttt{NaPS+GPMC}, and \texttt{KUT\_KMHT}, respectively solved 110, 105, and 99 benchmarks (Table~\ref{table:score}).

\begin{table}[t] 
  \caption{Score improvement by portfolios between DP and MC solvers}
  \label{table:coverage}
  \begin{center}
    \begin{tabular}{l|l|r|r}
      \toprule
        \multicolumn{2}{c|}{Portfolio: solver (approach)} & \# solved b'marks  & Improvement     \\
      \midrule
      \texttt{Drifters} (DP)                              & \texttt{NaPS+GPMC} (MC) & 122 & $+12$ \\
      \texttt{KUT\_KMHT} (DP)                             & \texttt{NaPS+GPMC} (MC) & 121 & $+16$ \\
      \texttt{Drifters} (DP)                              & \texttt{KUT\_KMHT} (DP) & 114 & $+4$  \\
      \bottomrule  
    \end{tabular} 
  \end{center} 
\end{table}

We investigate how much the scores could be improved by a portfolio solver that employs two of the three solvers;
the larger the score improvement, the more different characteristics the solvers have.
As shown in Table~\ref{table:coverage}, a portfolio of \texttt{Drifters} and \texttt{NaPS+GPMC} increases the score by 12 benchmarks, while a portfolio of DP solvers (\texttt{Drifters} and \texttt{KUT\_KMHT}) increases it just by 4 benchmarks.
Another portfolio of DP and MC (\texttt{NaPS+GPMC} and \texttt{KUT\_KMHT}) greatly increases the score as well, by 16 benchmarks.
These results show that the DP and MC approaches have different competencies, and that both are essential for the development of graph counting algorithms.

\section{Conclusions}
\label{sec:conclusion}

This paper minutely described the first GC competition, the ICGCA.
The authors thank all the contestants for their enthusiasm and their strong contributions. 
We are pleased that so many solvers performed so well on the challenging benchmarks.

Since no competition on GC has ever been held before, we conducted the preliminary experiments to select the benchmark set of reasonable hardness levels.
However, the winning solver solved more than 90\% of the benchmarks, so it could have been more challenging.

For \#P problems including GC, verifying the correctness of a given answer is not trivial, unlike NP problems.
Thus, the scores could be given with a range, resulting in unfixed rankings; future competitions on \#P problems must deal with this ambiguity.

In this competition, all the contestants submitted just a single version of their solvers, although in other competitions contestants often submitted multiple versions with different configurations.
This is because the final benchmark set was made available online in ICGCA and the contestants could adjust their configurations before submission.
The pros and cons of this decision are open to debate.

The main goal of the ICGCA was to provide a unified stage for the three independently developed approaches and to encourage their interaction.
We believe that this goal was accomplished to some extent.
Only the winning solver \texttt{TLDC} implements all three approaches, suggesting that each approach has its own potential.
Our analysis showed that both the DP and MC approaches have different strengths, so we expect them to fuel the future development of GC algorithms.
On the other hand, unfortunately, no detailed analysis of the BT approach was performed, because the \texttt{PCSSPC} solver, which is primarily based on the BT approach, failed to work properly.
However, since experiments by the \texttt{TLDC} developers showed that the most contributing approach in \texttt{TLDC} is BT, we also look forward to future contributions from this approach.

As the first competition on GC algorithms, we kept the competition simple.
In the future edition, it is worth considering to hold a main (sequential) track and a parallel track separately, possibly along with an approximate track.
It is also worth employing the PAR-2 scoring system to reflect the computation time in the ranking.
Anyone interested in the competition is welcome to contact us and join the discussion.



\section*{Acknowledgments}
This work was supported by JSPS KAKENHI Grant Numbers 20H05961, JP20H05963, and 20H05964. 


\bibliographystyle{unsrtnat}
\bibliography{ref}  






\end{document}